\documentclass[prb,preprint]{revtex4}

\pdfoutput=1

\usepackage{graphicx}

\begin{document}

\title{
\emph{Less} trouble with orbits: The semi-classical hydrogen problem in parabolic and spherical coordinates

\medskip }

\date{June 3, 2014} \bigskip

\author{Manfred Bucher \\}
\affiliation{\text{\textnormal{Physics Department, California State University,}} \textnormal{Fresno,}
\textnormal{Fresno, California 93740-8031} \\}

\begin{abstract}
Historically, the eccentricity of Sommerfeld orbits from quantization conditions in either parabolic or spherical coordinates was found to differ in almost all cases. To do the orbit comparison correctly, one must use \emph{amended} instead of traditional Sommerfeld orbits.
\end{abstract}

\maketitle

\section{INTRODUCTION}

There is no question that the orbit-based old quantum theory of the hydrogen atom by Bohr and Sommerfeld is incomplete.  While it accounts for the particle character of the electron, it completely ignores its wave character. One consequence is that the orbits depend on the choice of the coordinate system.
This was one of the reasons why historically the old quantum theory was dismissed.  Another reason concerned wrong values of orbital angular momentum.  The subsequently developed quantum mechanics of Heisenberg (matrix mechanics) and Schr\"{o}dinger (wave mechanics) resolved all these problems. In the meantime it has been found that the difficulty with orbital angular momentum originates with Sommerfeld's discarding of straight-line electron orbits through the nucleus, called ``Coulomb oscillators.'' \cite{1} When these are included (and circular Bohr orbits of principal quantum number $n$ consequently omitted) by the use of angular quantum number 

\begin{equation}
 l = 0, 1, ..., n-1,
\end{equation}

\noindent instead of $l = 1, 2, ..., n$,                               
along with a correction of both the orbit's semi-focal axis $f_{nl}$ (semilatus rectum) and angular momentum $|\textbf{L}_{nl}|$ in proportion

\begin{equation}
 f_{nl} \propto   |\mathbf{L}_{nl}|^2  \propto  l(l+1),
\end{equation}

\noindent instead of $\propto  l^{2}$,                                           
agreement of those orbit quantities with angular orbital momentum and average size of quantum-mechanical orbitals is achieved.\cite{1}

On the other hand, the non-uniqueness of Bohr-Sommerfeld orbits, that is, their dependence on the coordinate system, persists.  Historically, it first arose in the old quantum theory of the (linear) Stark effect---the splitting of spectral lines in an external electric field $E$---independently developed by Epstein and Schwarzschild.\cite{2} A great success at the time, it accounts for the experimental findings and agrees with subsequent quantum-mechanical results.  The theory obtains orbits for the electron of a hydrogen atom when quantization conditions in parabolic coordinates are imposed.  However, in the absence of an electric field, $E = 0$, these orbits disagree with the traditional Sommerfeld orbits of the $H$ atom, quantized in spherical coordinates---hence the conflict of non-uniqueness. 

\section{FORMALISM}

Geometrically, two planar ellipses are equal if they agree in the length of two of their axes, say the length of major axis, $2a$, and of focal axis (latus rectum), $2f$.  Equivalently, two ellipses of the same length of major axis are equal if they agree in their eccentricity,

\begin{equation}
\varepsilon = \sqrt{\frac{a - f}{a}}.
\end{equation}


\noindent By the Bohr-Sommerfeld theory of the hydrogen atom the semi-major axis of an elliptical orbit is
\begin{equation}
a_{nl} = n^{2} r_{B}                                                              
\end{equation}

\noindent where $r_{B}$ is the Bohr radius and $n$ is the principal quantum number.  
In terms of spherical quantization conditions, $n$ is the sum of the radial and angular quantum numbers,
\begin{equation}
n = n_{r} + n_{\theta} + n_{\varphi} = n_{r} + l ,                                         
\end{equation}

\noindent aptly called ``quantum sum'' by Sommerfeld.
Expressed in terms of parabolic quantum numbers $n_\xi$ and $n_\eta$, and the azimuthal, or magnetic, quantum number $n_\varphi = m$, the quantum sum is
\begin{equation}
n = n_{\xi} + n_{\eta} + |m|.                                        
\end{equation}

\noindent For a vanishing external electric field, $E = 0$, the eccentricity of an electron orbit from quantization conditions in parabolic coordinates is given by
\begin{equation}
\varepsilon_{\xi \eta |m|} = \frac{1}{n}\Bigg[\sqrt{n_{\xi}(n_{\xi}+ |m|)}+\sqrt{n_{\eta}(n_{\eta}+ |m|)}\Bigg].
\end{equation}

When this is compared with the eccentricity of a traditional ($n,l,m$) Sommerfeld orbit with quantization conditions in spherical coordinates,
\begin{equation}
\varepsilon_{nlm}=\frac{1}{n} \sqrt{n^2 - l^2},
\end{equation}

\noindent disagreement is found in all cases except the circular Bohr orbits ($\varepsilon = 0$).\cite{2}  However, since the traditional Sommerfeld orbits suffer from the defects mentioned above, one must use in this comparison the correct sequence and width of Sommerfeld orbits according to Eqs. (1) and (2).  The eccentricity of such \emph{amended} Sommerfeld orbits with spherical quantization conditions, denoted  $\langle n l m \rangle$, is then
\begin{equation}
\varepsilon_{nlm}=\frac{1}{n} \sqrt{n^2 - l(l+1)},
\end{equation}

\noindent instead of Eq. (8).
This provides agreement of orbit eccentricity, $\varepsilon_{\xi \eta |m|} = \varepsilon_{nlm}$,
in the cases of extreme $|m|$ and close values in all other cases.  It also reveals relationships between orbits from quantization conditions in both coordinate systems for common quantum numbers $n$ and $|m|$.

\section{RESULTS}

Table I gives calculated eccentricities $\varepsilon_{\xi \eta |m|}$ and $\varepsilon_{n l m}$ for principal quantum numbers $n = 1, 2, 3$.  By Eq. (1), circular Bohr orbits, $l = n$ and $\varepsilon = 0$, are excluded from the amended Sommerfeld model (cases 2, 5, 11).  The ground state, $\{1 0 0\}$ or $\langle 1 0 0 \rangle$, is a straight-line orbit through the nucleus, called Coulomb oscillator, with eccentricity $\varepsilon_{\xi \eta |m|} = \varepsilon_{n l m} = 1$, in agreement for quantization in both coordinate systems (case 1).  So are higher Coulomb oscillators (cases 3 and 6).  Agreement of eccentricity, $\varepsilon_{\xi \eta |m|} = \varepsilon_{n l m}$, also holds when, for a given $n$, the magnetic quantum number takes on the maximum allowed value, 
$|m| = l = n-1$ (cases 1, 4, 10).  Disagreement is found for the cases 7 and 9 whose eccentricity $\varepsilon_{\xi \eta |m|}\{1 1 1\} = \frac{1}{3}\sqrt{8} = 0.94$ and $\varepsilon_{\xi \eta |m|}\{2 0 1\} = \frac{1}{3}\sqrt{6} = 0.82$ flank the amended Sommerfeld value  $\varepsilon_{n lm}\langle 3 1 m \rangle = \frac{1}{3}\sqrt{7} = 0.88$ by  $\pm 0.06$.  

The same pattern as in Table I can be seen in Table II which lists the eccentricities of orbits with quantum sum $n = 4$.  Agreement between the parabolic and spherical expressions, Eqs. (7) and (9), holds for Coulomb oscillators, $|m| = 0$ (case 12), and for maximum $|m|$, here $|m| = 3$ (case 19).
Otherwise the values of the parabolic expression of eccentricity (cases 13 and 15 or 16 and 18) flank the eccentricity values of the related Sommerfeld ellipses, $|m|=l$ (cases 14 or 17, respectively).
The inspection of Tables I and II then raises the question, why is there agreement in the extreme cases of minimum and maximum  $|m|$ and disagreement otherwise?

\section{DISCUSSION}
By Eq. (9) the eccentricity of an amended Sommerfeld orbit depends only on the principal and angular quantum numbers, $n$ and $l$, not on the magnetic quantum number $m$.  The latter determines different orientations of a given $\langle n l m \rangle$ Sommerfeld ellipse 
(space quantization) for

\pagebreak
\includegraphics[width=6in]{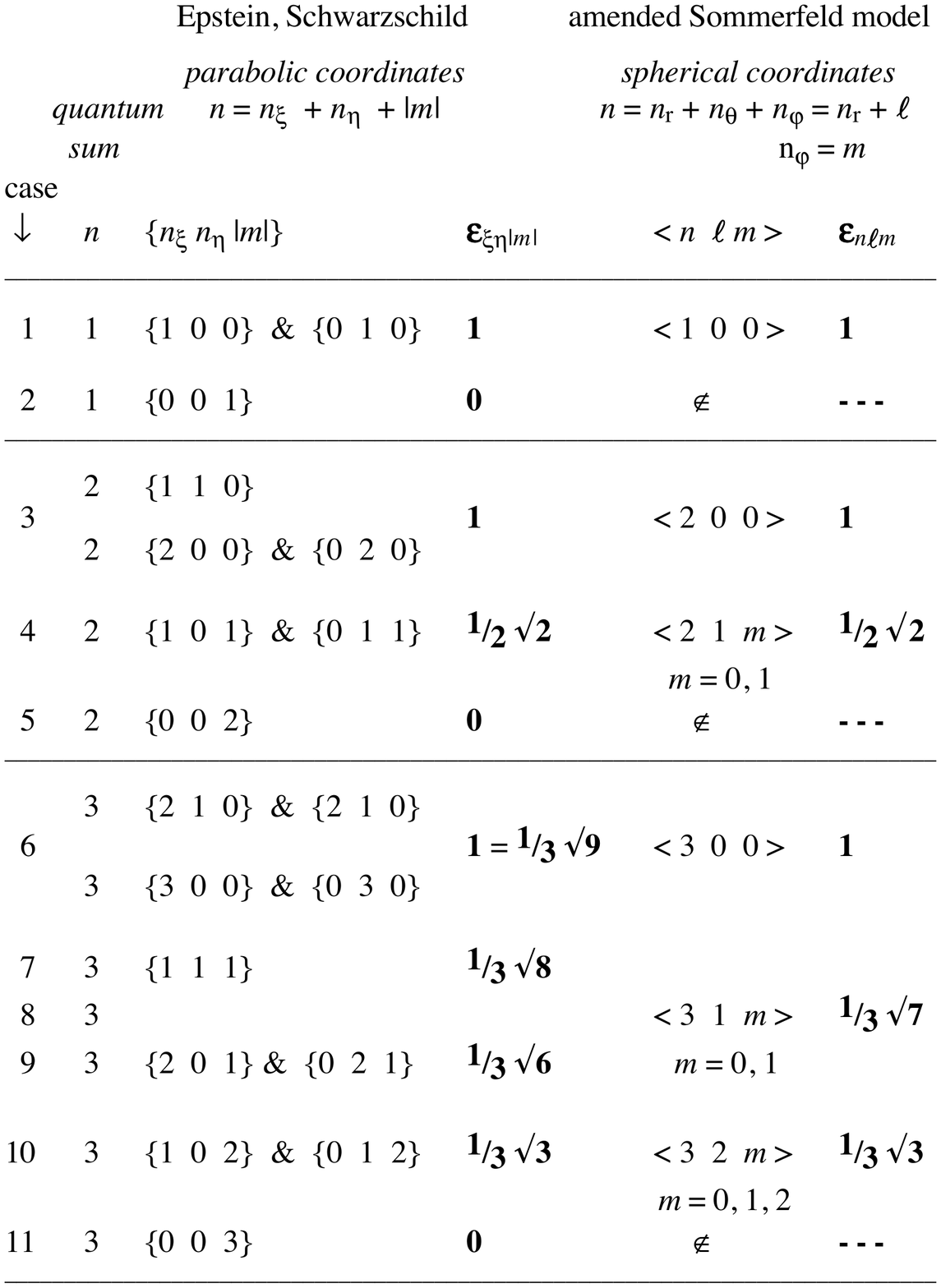}
\noindent Table I. Eccentricities of Sommerfeld orbits for principal quantum number $n = 1, 2, 3$ with parabolic quantization conditions $\{n_{\xi} n_{\eta} |m|\}$ (left side) and spherical quantization conditions $\langle nlm\rangle$ (right side).  The eccentricity $\varepsilon_{\xi \eta |m|}$ was calculated with Eq. (7) and $\varepsilon_{nlm}$ with Eq. (9). The symbol $\notin$ here denotes non-existence.
\pagebreak

\includegraphics[width=6in]{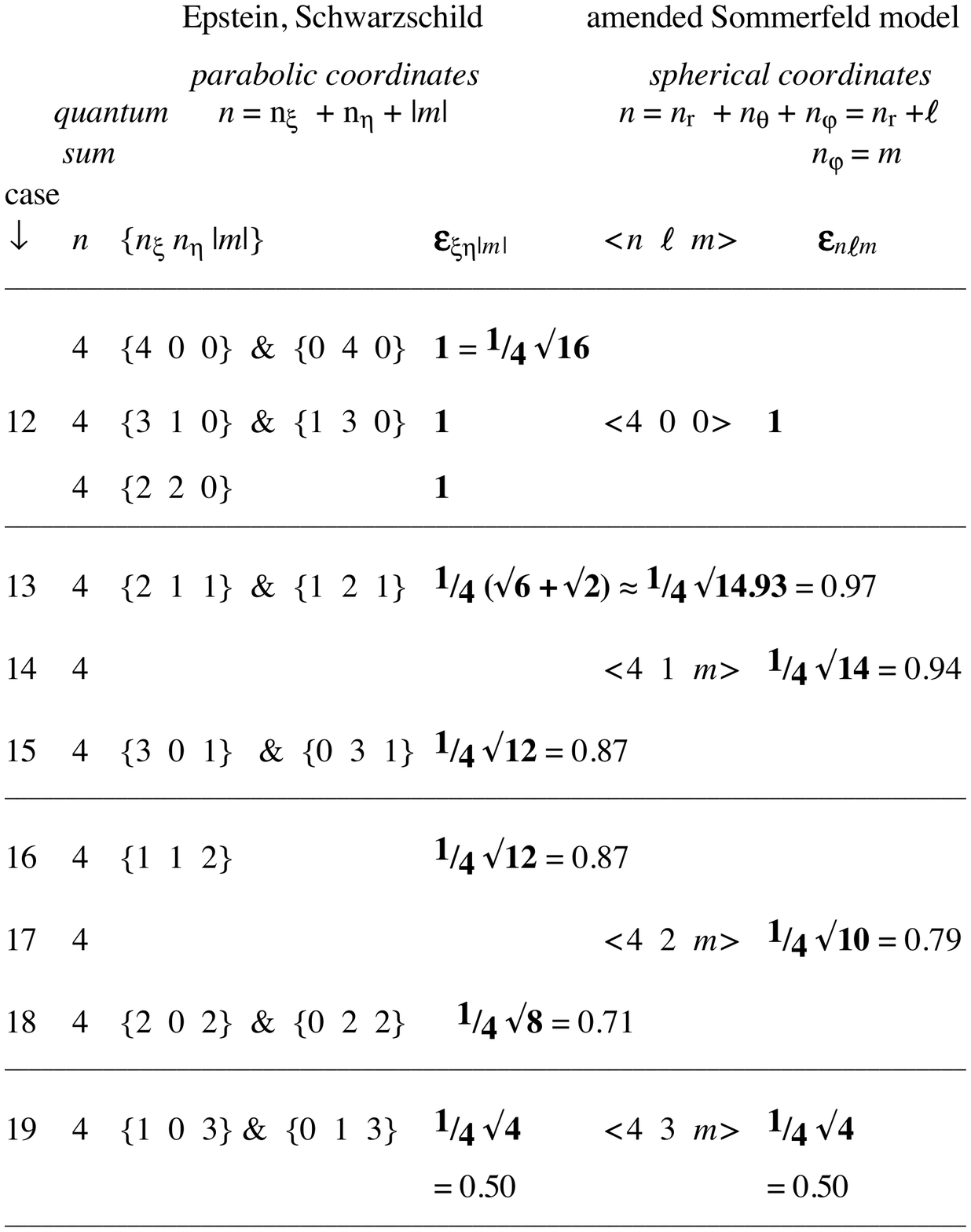}
\noindent Table II. Eccentricities of Sommerfeld orbits for principal quantum number $n = 4$ with parabolic quantization conditions $\{n_{\xi} n_{\eta} |m|\}$ (left side) and spherical quantization conditions $\langle nlm\rangle$ (right side).  The eccentricity $\varepsilon_{\xi \eta |m|}$ was calculated with Eq. (7) and $\varepsilon_{nlm}$ with Eq. (9).
\pagebreak

\begin{equation}
|m| = 0, 1, ..., l.
\end{equation}

\noindent An apt expression for $m$ would be ``\emph{lean} quantum number'', as it affects the lean angle,
\begin{equation}
\theta = arcsin \frac{|m|}{\sqrt{l(l+1)]}}, 
\end{equation}

\noindent by which the minor axis of a Sommerfeld ellipse of given major axis, $2a = 2n^2 r_{B}$, and focal axis, $2f = 2l(l+1) r_{B}$, leans away from the coordinate $z$ axis.
This aligns the minor axis of an $\langle n l 0 \rangle$ orbit parallel to the $z$ axis, $\theta_{min} = 0$, and gives the largest lean angle of an $\langle n l |m| \rangle$ orbit, 
\begin{equation}
\theta_{max} = arcsin\frac{n-1}{\sqrt{n(n-1)}} < \frac{\pi}{2},
\end{equation}
when 
\begin{equation}
|m| = l = n-1.
\end{equation}

\noindent The latter case is the only instance where the angular quantum number $l$ of the orbit is fully known from the magnetic quantum number $|m|$ in the spherical quantization conditions
$\langle nlm \rangle$ under the constraint of Eq. (10).

Different constraints on the quantum numbers hold for quantization in parabolic coordinates.  
An analysis of them is beyond the scope of this paper. However, four patterns are noticeable:
 (1) If $m=0$ in $\{n_{\xi},n_{\eta},|m|\}$, then $\varepsilon_{\xi \eta |m|}=0$ and the orbit is a Coulomb oscillator (cases 1, 3, 6, 12).
\newline (2) If $|m|$ is maximal in $\{n_{\xi},n_{\eta},|m|\}$, that is $\{1,0,n-1\}$ or  $\{0,1,n-1\}$, then one can assign a ``good'' (integer) angular quantum number $l=n-1$ to the orbit, in agreement with its role in an $\langle n,l,m\rangle$ orbit (cases 1, 4, 10, 19). The eccentricity then is
\begin{equation}
\varepsilon_{\xi \eta |m|}\{1,0,n-1\} = \varepsilon_{nlm}\langle n,n-1,m\rangle =
\frac{1}{n} \sqrt{n},
\end{equation} 

\noindent and the angular momentum $\textbf{L}$ of the $\{n_{\xi},n_{\eta},|m|\}$ orbit is solely determined by $|m|$.
\newline (3) Orbits $\{1,1,n-2\}$ have a fractional value of $l$, but such that $\sqrt{l(l+1)}=n-2=|m|$ (cases 3, 7, 16). Again, their angular momentum $\textbf{L}$ is solely determined by $|m|$.
\newline (4) In other cases (9, 15, 13) not only $|m|$ but the parabolic quantum numbers $n_{\xi}$ and $n_{\eta}$ too, contribute to the angular momentum $\textbf{L}$ via a fractional angular quantum number $l$ with $l(l+1)=3$, $=2$, and $\simeq 1$, respectively.
The contributing role of parabolic quantum numbers can be seen in case 15, achieving the same orbit eccentricity $\varepsilon_{\xi \eta |m|}$ as in case 16.

\pagebreak
\includegraphics[width=6in]{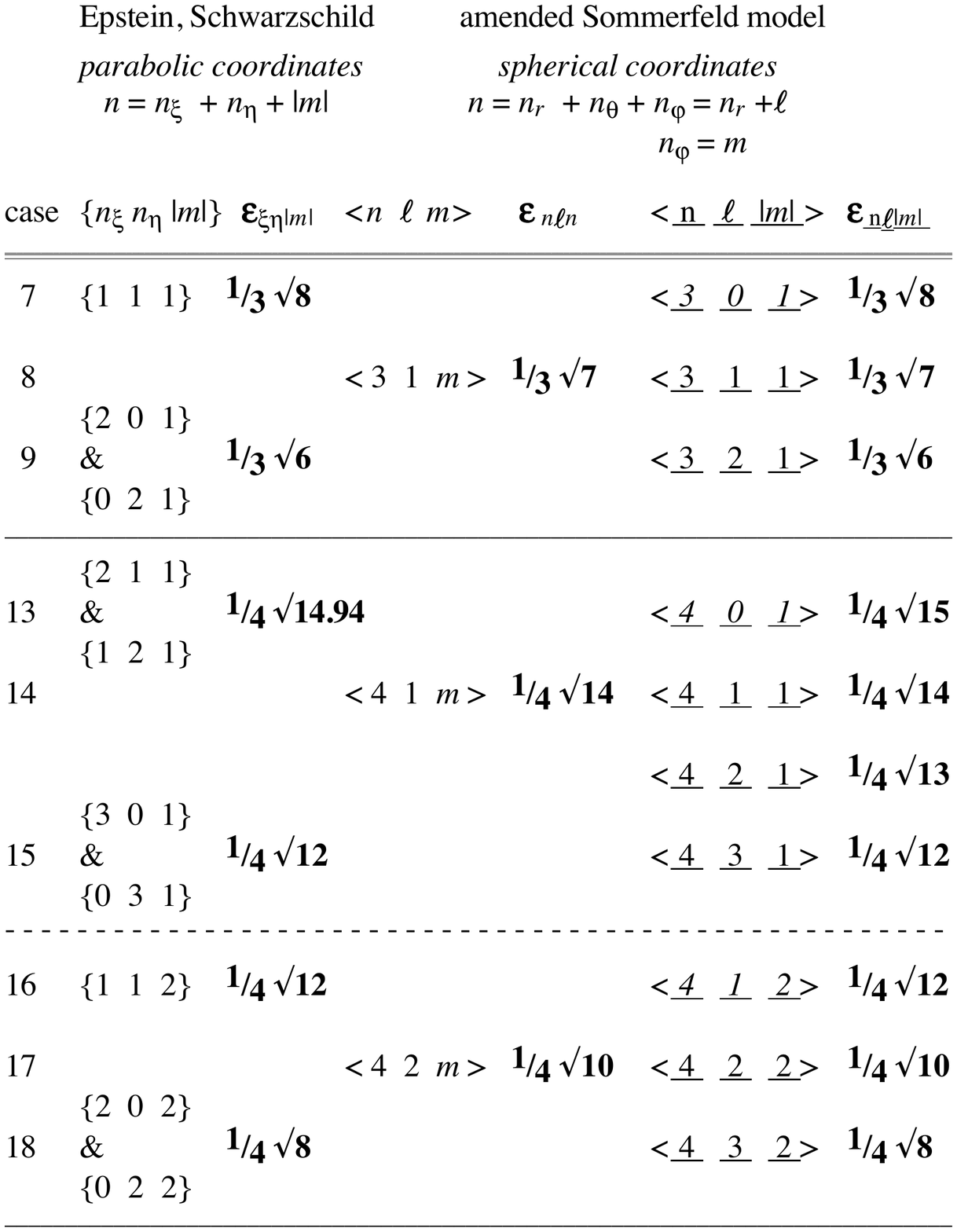}
\noindent Table III. Eccentricities of Sommerfeld orbits for the cases from Table I and II where disagreement is found, with parabolic quantization conditions [left side, Eq. (7)] and spherical quantization conditions [center columns, Eq. (9)].  The right side shows eccentricities of \emph{projected} orbits (denoted by underline) obtained with Eq. (15).
\pagebreak

\section{VISUALIZATION}
Imagine the projection of a leaning $\langle n l |m|\rangle$ Sommerfeld orbit onto the $xy$ plane. The minor axis of the \emph{projected} ellipse, denoted by underline as $\langle \underline{n} \underline{l} |\underline{m}|\rangle$, is shortened in proportion of $m/l$.\cite{3}  Combined with Eq. (9) this renders its eccentricity as
\begin{equation}
\varepsilon_{\underline{n} \underline{l} |\underline{m}|} = \frac{1}{n} \sqrt{n^2 - |m|(l+1)}.
\end{equation}

\noindent The expression agrees with the eccentricity in parabolic quantization, Eq. (7), for \emph{maximum} $|m|$, Eq. (13), as obtained in Eq. (14).

\noindent For the \emph{minimum} magnetic quantum number, $|m|=0$, Eq. (15) gives
\begin{equation}
\varepsilon_{\xi \eta |m|}\{n_{\xi},n_{\eta},0\} =
\varepsilon_{\underline{n} \underline{l} |\underline{m}|}\langle n, l, 0\rangle = 1.
\end{equation}

\noindent In this instance the $\langle nlm \rangle$ Sommerfeld ellipses don't lean, so their projection onto the $xy$ plane gives line ellipses of $\varepsilon=1$, as expressed in Eq. (16), which explains those values in Tables I and II.

\noindent In the remaining cases, for orbits with common $n$ and
\begin{equation}
|m| = l,
\end{equation}
\noindent agreement of $\varepsilon_{\xi \eta |m|}$ and $\varepsilon_{nlm}$ from Eqs. (7) and (9) is close but not exact, as Table III shows in the left and center columns. Instead, the eccentricity of the leaning $\langle nlm \rangle$ orbits (center column) agrees with that of their projected ellipses, 
$\langle \underline{n} \underline{l} |\underline{m}| \rangle$ (right column),
\begin{equation}
\varepsilon_{nlm} = \varepsilon_{\underline{n} \underline{l} |\underline{m}|},
\end{equation}

\noindent cases 8, 14, and 17, where the lean angle of a given $\langle nlm \rangle$ orbit is maximal, Eq. (17).

Next we compare the eccentricity of orbits with parabolic quantization conditions $\{n_{\xi}n_{\eta}|m|\}$ (left column) with that of projected ellipses $\langle \underline{n} \underline{l} |\underline{m}| \rangle$ (right column) for common $n$ and $|m|$. If
\begin{equation}
|m| < \underline{l}
\end{equation}

\noindent (cases 9, 15 and 18), agreement of eccentricity $\varepsilon_{\xi \eta |m|}$ from Eq. (7) can be found with \emph{one} of the $\varepsilon_{\underline{n} \underline{l} |\underline{m}|}$ solutions of Eq. (15).

\noindent In the opposite situation,
\begin{equation}
|m| > \underline{l}
\end{equation}

\noindent Eq. (15) still provides exact solutions,  $\varepsilon_{\xi \eta |m|} = \varepsilon_{\underline{n} \underline{l} |\underline{m}|}$, in two cases (7 and 16).  The only exception is case 13 which is numerically slightly off, $\varepsilon_{\xi \eta |m|} \simeq \varepsilon_{\underline{n} \underline{l} |\underline{m}|}$. The three cases are highlighted by underlined and italic notation, 
$\langle\underline{\textit{n}} \underline{\textit{l}} |\underline{\textit{m}}|\rangle$, on the right side of Table III.
Note that Eq. (20) does \emph{not} refer to the angular quantum number $l$ from quantization in spherical coordinates, as $|m| > l$ would violate the constraint of Eq. (10).  Instead Eq. (15), with (underlined) $\underline{l}$ from Eq. (20), expresses the eccentricity of orbits $\{111\}$, $\{112\}$ and $\{211\}$ which obey constraints of quantum numbers in parabolic quantization but not those of spherical quantization.

\section{CONCLUSION}

The historical comparison of orbit eccentricity, based on quantization conditions in parabolic coordinates and, respectively, spherical coordinates for traditional Sommerfeld orbits, gives disagreement in all cases except for the circular Bohr orbits. This suggests a large dissimilarity between the two sets of orbits.  When instead the comparison is done with the amended Sommerfeld model, the situation drastically changes:  The previous cases of agreement disappear with the rejection of circular orbits whereas the cases of previous disagreement become exact or close. 
The reason for only approximate agreement of some eccentricities originates with different constraints on quantum numbers in different coordinate systems.
All orbits with parabolic quantization conditions can be visualized as projected ellipses from leaning $\langle nlm \rangle$ Sommerfeld orbits onto the $xy$ plane.

The trouble with orbits continues---but less so than previously thought.

\bigskip
\centerline{ \textbf{ACKNOWLEDGMENT}}

\noindent I thank Duane Siemens for valuable discussions and critique.


\begin{thebibliography}{3}

\bibitem{1} M. Bucher, ``Rise and premature fall of the old quantum theory,'' arXiv:0802.1366v1

\bibitem{2} A. Duncan and M. Janssen, ``The trouble with orbits: The Stark effect in the old and the new quantum theory,'' arXiv:1404.5333v1

\bibitem{3} The choice of the projection ratio $m/l$ is \textit{ad hoc}, justified only by its numerical result and the potential of visualization. It differs from the ratio $m/\sqrt{l(l+1)}$ for the lean angle, Eq. (11).

\noindent 

\end{thebibliography}
\end{document}